\documentclass[conference]{vgtc}
\ifpdf
  \pdfoutput=1\relax                   
  \usepackage{graphicx}                
  \DeclareGraphicsExtensions{.pdf,.png,.jpg,.jpeg} 
\else
  \usepackage{graphicx}                
  \DeclareGraphicsExtensions{.eps}     
\fi%

\graphicspath{{figures/}{pictures/}{images/}{./}} 
\usepackage{enumitem}
\usepackage{microtype}                 
\PassOptionsToPackage{warn}{textcomp}  
\usepackage{textcomp}                  
\usepackage{mathptmx}                  
\usepackage{times}                     
\usepackage{cite}                      
\usepackage{tabu}                      
\usepackage{booktabs}                  
\usepackage{caption}
\usepackage[normalem]{ulem}
\usepackage{xspace}
\newcommand {\circled}[1]{\textcircled{\small {#1}}}
\newcommand {\mycircled}[1]{\textcircled{\small {#1}}}

\onlineid{1178}

\vgtccategory{Research}
\vgtcpapertype{System or Tool}

\newcommand{\app}{AdViCE}

\title{\textit{\app}: Aggregated Visual Counterfactual Explanations for Machine Learning Model Validation}

\author{Oscar Gomez\thanks{Both authors contributed equally to this research\newline E-mails: [oscar.gomez, steffen.holter, junyuan, enrico.bertini]@nyu.edu}, Steffen Holter\footnotemark[1], Jun Yuan and Enrico Bertini}

\shortauthortitle{Gomez \MakeLowercase{\textit{et al.}}: \textit{\app}: Aggregated Visual Counterfactual Explanations for Machine Learning Model Validation}

\abstract{Rapid improvements in the performance of machine learning models have pushed them to the forefront of data-driven decision-making. Meanwhile, the increased integration of these models into various application domains has further highlighted the need for greater interpretability and transparency. To identify problems such as bias, overfitting, and incorrect correlations, data scientists require tools that explain the mechanisms with which these model decisions are made. In this paper we introduce \textit{\app}, a visual analytics tool that aims to guide users in black-box model debugging and validation. The solution rests on two main visual user interface innovations: (1) an interactive visualization design that enables the comparison of decisions on user-defined data subsets; (2) an algorithm and visual design to compute and visualize counterfactual explanations - explanations that depict model outcomes when data features are perturbed from their original values. We provide a demonstration of the tool through a use case that showcases the capabilities and potential limitations of the proposed approach.
} 


\teaser{
  \centering
  \includegraphics[width=0.85\linewidth]{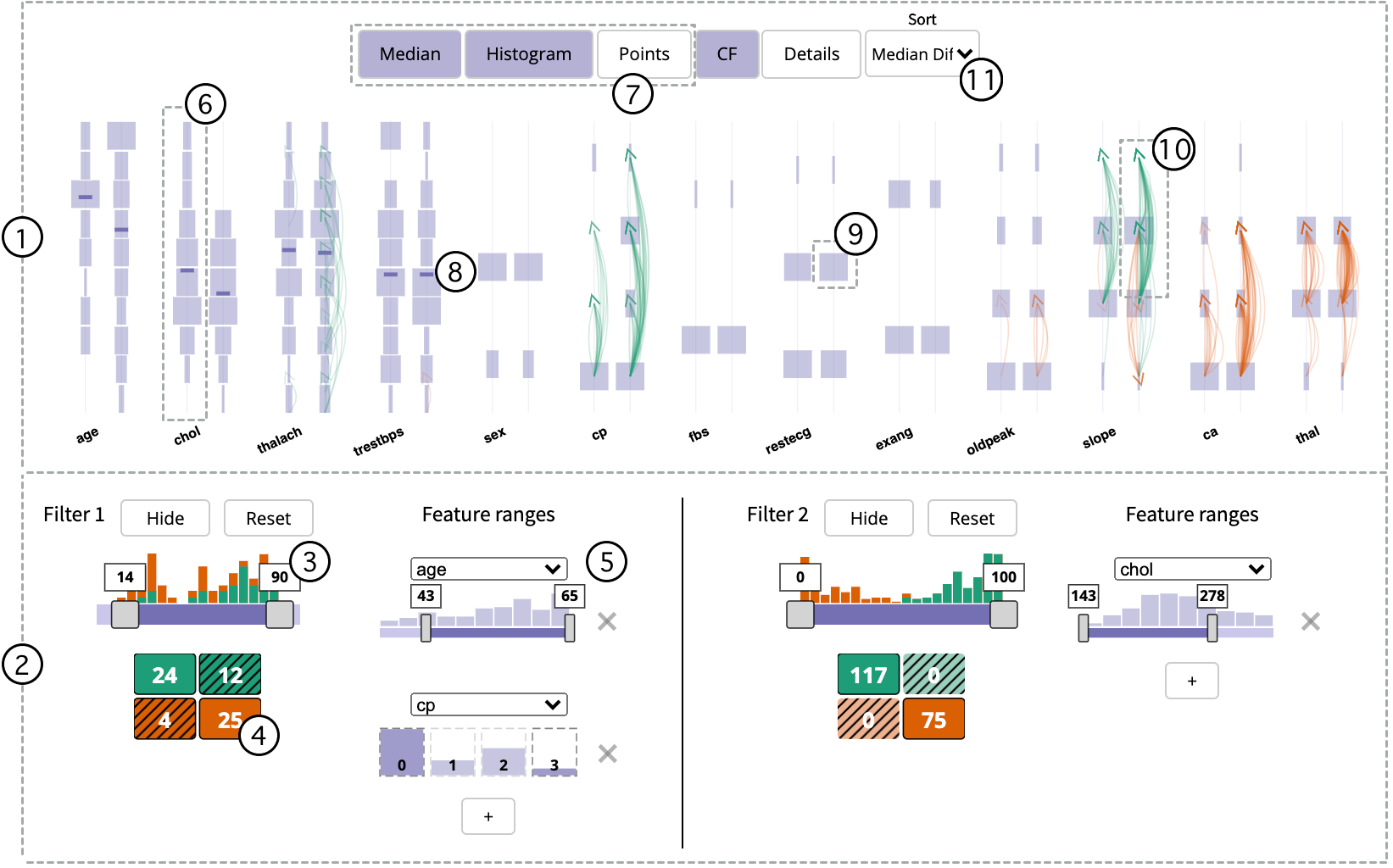}
   \vspace*{-.5cm}
   \caption{\textbf{Complete visual interface for AdViCE.} \circled{1} main visualization, \circled{2} filtering section, \circled{3} model prediction percentage, \circled{4} confusion matrix, \circled{5} feature range selector, \circled{6} feature sub-column, \circled{7} information toggles, \circled{8} median value, \circled{9} histogram bin, \circled{10} set of counterfactual explanations, \circled{11} sorting function}
    \vspace*{-.5cm}
    \label{fig:interface}
}

\teaser{
  \centering
  \includegraphics[width=0.80\linewidth]{images/main-vis.png}
  \caption{\textbf{Complete visual interface for AdViCE.} \mycircled{1} main visualization, \mycircled{2} filtering section, \mycircled{3} model prediction range selector, \mycircled{4} confusion matrix, \mycircled{5} feature range selector, \mycircled{6} feature sub-column, \mycircled{7} information toggles, \mycircled{8} median value, \mycircled{9} histogram bin, \mycircled{10} set of counterfactual explanations, \mycircled{11} sorting method}
  \label{fig:interface}
}

\begin{document}


\firstsection{Introduction}

\maketitle



Model interpretability is one of the key challenges in the field of artificial intelligence (AI). Machine learning models are usually limited by their inability to explain their reasoning to human users. However, understanding the rationale behind decisions is a clear necessity to ensure trust, fairness and accountability \cite{doshi2017towards, miller2017explainable}. 
In domains such as healthcare, where associated risks are greater, more than a single accuracy measure is required for the models to be integrated into the decision making process. Even for well fitted models the absence of hidden flaws such as biases and false generalizations cannot be ensured without model transparency. Especially, \textit{model developers} – individuals responsible for developing and testing models – need to be able to verify that the elicited predictions coincide with the expected behavior within the application domain. \looseness=-1


In this paper, we introduce \textit{\app}\footnote{https://github.com/5teffen/AdViCE} – \textit{Aggregated Visual Counterfactual Explanations}, a novel explainable machine learning visual analytics tool. We show that by aggregating and comparing groups of counterfactual (CF) explanations – minimal sets of changes needed to change the model’s output – the tool can help users understand how models behave. As such, we see our tool catering primarily to data scientists dealing with model validation and debugging. In particular, \textit{\app} is built to help \textit{model developers} increase confidence in the reliability and validity of the models they construct.

Analysis with \textit{\app} is driven by the juxtaposition of what the model does right and what the model does wrong. By allowing the user to create custom data subsets which act as ``test cases'', the model's performance can be evaluated against the expected performance envisioned by the user. 
Furthermore, it is our belief that by allowing data scientists to freely explore, compare and visualise custom subsets of data points they can discover unique insights into the operation of the model that extend beyond their initial expectations and knowledge. Thus, \textit{\app} provides both a way to create hypotheses and to verify them. \looseness=-1


The explanation of model behavior relies on the introduction of multiple visual counterfactual explanations. Our tool makes use of a fully model-agnostic heuristic algorithm for calculating these counterfactuals. By aggregating and visualizing the counterfactual changes for groups of individual data points, global interpretability can be inferred. \textit{\app} also provides a way to contextualize these explanations through direct data comparison and custom filtering. Our contributions are as follows: \looseness=-1

 
\begin{enumerate}[wide ,nosep]
    \item[\textbf{1. AdViCE interface –}] A tool for visualizing sets of contextualized counterfactuals and comparing user-defined data subsets.
    \item[\textbf{2. Counterfactual algorithm –}] A model-agnostic heuristic method to generate counterfactuals for tabular data.
    \item[\textbf{3. Model debugging work-flow –}] A formalization of a multi-step workflow to facilitate model exploration and verification.
\end{enumerate}

\section{Background And Related Work} 







The need for machine learning (ML) model interpretability has been recognized by researchers and practitioners ~\cite{miller2019explanation,doran2017does,gunning2019xai,lipton2018mythos}. A recent interview with industry experts revealed that the goals of ML model interpretation include model validation, knowledge discovery, and gaining trust ~\cite{hong2020human}. The design of \textit{\app} falls into the category of assisting data scientists with ML model validation.\looseness=-1

Previous works~\cite{biran2017explanation,doshi2017towards,lipton2016mythos,poursabzi2018manipulating,molnar2019,guidotti2019survey,adadi2018peeking,carvalho2019machine} provide an overview of methods, opportunities and challenges in the area of machine learning model explanation. These methods can be  broadly categorized into two classes: \textit{white-box} and \textit{black-box}. White-box approaches aim at explaining the internal structure of a model; while black-box methods are applied to explain a ML model's decisions without exposing the internal operation of the model. As \textit{\app} has the purpose of providing explanations for a range of different models, our focus will be on model-agnostic techniques. \looseness=-1

Black-box approaches to ML explanation are usually categorized as \textit{local} or \textit{global}, according to whether the explanation is used for a specific data instance or to describe the behavior of the model as a whole. Popular examples of local approaches include LIME~\cite{ribeiro2016should} and SHAP~\cite{lundberg2017unified}, methods which assign weights to compare the importance of each feature in a specific decision. Anchors~\cite{ribeiro2018anchors} explains a single prediction using an if-then rule and LORE~\cite{guidotti2018local} provides an if-then rule for each decision while supplementing it with a counterfactual explanation. To achieve understanding of model behavior at a global scale, local explanations need to be organized and visualized in a way that provides insight on sets of multiple instances. Researchers have also explored the idea of establishing global explanations by training rule based models using the output of a model they want to simulate ~\cite{craven1996extracting,sanchez2015towards}.

Counterfactual explanations are a type of local explanation that can be defined as the minimal set of changes needed in the feature values to flip the prediction of that instance. For example, finding the smallest feature perturbation that would change the prediction of a loan application from rejected to approved. Multiple approaches have been developed with regards to the generation of counterfactual explanations. Wachter \textit{et al.}~\cite{wachter2017counterfactual} describes a general framework for counterfactual generation using stochastic optimization; and Ustun \textit{et al.}~\cite{ustun2019actionable} present an approach specific to linear classifiers. A recent work, DiCE~\cite{mothilal2020explaining}, generates a diverse set of counterfactual explanations for classifiers. In \textit{\app}, we build upon and further develop the counterfactual explanation algorithm proposed by Martens {et al.}~\cite{martens2013explaining}, extending it for use on tabular data (see section~\ref{sec:algorithm}). \looseness=-1

Visual analytics has been successfully applied for model explainability in recent years. However, there are limited works proposed for visualizing counterfactual explanations specifically. Recently, the What-If Tool (WIT)~\cite{wexler2019if} and Seq2SeqVis~\cite{strobelt2018s} have been suggested to provide support and answer what-if questions. Similarly, Rivelo~\cite{tamagnini2017interpreting} visualizes counterfactual explanations for text classification models. Krause \textit{et al.}~\cite{krause2017workflow} introduced a workflow that is based on instance-level counterfactual explanations for sparse binary data. However, their solution uses an algorithm ~\cite{martens2013explaining} and visual design that does not work with tabular data. Our work extends this algorithm to situations pertaining to continuous numerical and categorical data and proposes new visualization solutions for their exploration. ViCE \cite{gomez2020vice} has been proposed to visualize counterfactual explanations, however, this system only supports visualizing a single counterfactual explanation at a time with no way to display multiple explanations at once, and caters to \textit{end-users} of the models rather than \textit{model developers}. Closest to our work is the DECE tool \cite{cheng2020dece}, which also allows for the visualization of counterfactuals. However, its focus is on bias detection and visualises detailed counterfactuals only at instance level. Furthermore, the aggregation of explanations is applied one feature at a time and thus fails to explore relationships between multiple features. Through \textit{\app}, we propose a new design centered around the flexible comparison of aggregated sets of multi-feature counterfactuals. \looseness=-1

\section{Design Rationale} 

The main goal of this visual analytics tool is to provide a focused solution to assist in model understanding and debugging. The interface was developed using an iterative design process throughout which we attempted to identify how \textit{model developers} attempt to ``make sense'' of the models they develop. We concluded that this process generally involves two types of approaches. Firstly, \textbf{exploration} deals with identifying what the model does well and what it does \textit{not} do well, that is, the errors the model makes. This step is characterized by inferences where one first identifies erroneous cases and issues with the model and then works towards finding the reason and devising an explanation for the problem identified. On the other hand, \textbf{verification} is the process of checking how the model behaves with specific test cases that the user has in mind and wants to explore. This approach is characterized by first ideating and creating a hypothesis and then verifying the extent to which such a hypothesis is confirmed by the model behavior.\looseness=-1

Domain experts have cited creating test cases, determining feature importance, and understanding feature combinations as some of the key issues in model validation \cite{hong2020human,liao2020questioning}. With this in mind, we propose a multi-step workflow ideated to facilitate and guide the user in performing the two tasks above.


\begin{itemize}[nosep]
    \item[\textbf{S1}] \textbf{Filter}: 
    The filtering step is necessary to focus on instances that have some characteristics of interest. Creating these test cases is key to gaining confidence in the model and ensuring expected performance ~\cite{hong2020human}. A test case might be in the form of an individual data point or a larger data subset. 
    
    
    \item[\textbf{S2}] \textbf{Characterize and Compare}: After filtering the data to focus on a specific test case of interest, the next step requires contextualizing it with respect to the rest of the data. For instance, one should be able to identify that the subset has a higher or lower mean for a given set of features compared to the rest of the data or that the dispersion is narrower or wider. However, characterizing the distribution of an individual subset by itself is less insightful than comparing multiple subsets. Therefore, comparison becomes crucial, as it permits the user to directly observe the differences and focus their attention on the key aspects of the subsets. 
    
    
    \item[\textbf{S3}] \textbf{Explain}: Comparing distributions is useful for identifying trends and generating initial hypotheses about model behavior. However, this is of limited use if one cannot verify that a given rationale is actually reflective of the models' behavior. This problem can be solved by directly accessing the decision space of the model and using counterfactuals to gain insights about the role individual features play in the model's predictions.

\end{itemize}

Our proposed tool \textit{AdViCE} has been designed to support this workflow. The following section will explain the details of our solution and how we assist in carrying out these three steps.


\section{AdViCE} 

\subsection{Algorithm}
\label{sec:algorithm}

Our implementation for generating counterfactual explanations is a generalization of the algorithm proposed by Martens et al. \cite{martens2013explaining}. Originally meant for binary data for text classification, we extend it for use with continuous and categorical tabular data. The algorithm is a greedy heuristic which maximizes the amount of change in the model's prediction at each step. First, the data is discretized by fitting a Gaussian on each of the continuous features and splitting these values into \textit{n} bins such that the middle \textit{n-2} capture four standard deviations from the mean, and the extreme bins capture data points beyond this threshold. The algorithm greedily moves feature values across the bins (for continuous features) and across categories (for categorical features) until the predicted class is changed, or until pre-defined constraints (no more than $w$ features are changed in a single explanation and no feature value is moved across more than $l$ bins) are reached. This method ensures that the changes are at the same time interpretable and feasible, which is crucial for user-friendly explanations. \cite{miller2019explanation}. \looseness=-1

While more sophisticated methods for generating counterfactual explanations have been proposed \cite{mothilal2020explaining, mahajan2019preserving}, we use the above algorithm for several reasons. The heuristic proposed is easy to implement, fast to run, and is fully model-agnostic. Other more complex methods require that the models be differentiable, as they rely on gradient computations, and therefore limit the variety of models that a user of \textit{\app} could analyze. In addition, its simplicity allows users to easily configure additional constraints that the counterfactuals must obey. Even though we use the above algorithm in our implementation, it must be noted that the counterfactual generation step can be fully decoupled from the visualization stage, and therefore any other readily available method for generating counterfactuals can be used with \textit{\app}. Similarly, the counterfactual generation algorithm is independently available and can be used in other workflows. \looseness=-1

\subsection{Visual Interface}
\label{sec:visual-interface}

Fig.~\ref{fig:interface} shows the complete \textit{\app} visual interface. In order to effectively describe the design choices and operation of the tool we use the \textit{Heart Disease dataset} \cite{janosi1988heart} as a running example. Similarly, to simulate a black-box model, we train a support vector machine (SVM) classifier using a linear kernel. This acts as the target model being debugged and validated with the tool. \looseness=-1


\textit{\app} is designed according to the three step workflow proposed in the previous section. Emphasis is put on effective filtering, facilitating visual comparisons between data subsets, and inferring explanations. The steps of the workflow are also directly referenced in the deconstruction of the tool's interface (e.g., \textbf{S2)}. The tool itself is split into two main components: the filtering section (Fig.~\ref{fig:interface} \textcircled{2}) - used to select the custom subsets - and the comparative visualisation section (Fig.~\ref{fig:interface} \textcircled{1}) - used to visually compare and explain the results. \looseness=-1

The filtering section presents the user with two sets of dynamic filters, each corresponding to one half in the central visualisation. These toggles can be manipulated independently or in combination to generate custom test cases of select data points that the user is interested in \textbf{(S1)}. By default the tool initializes with two filter sets to emphasise the comparative aspect, however, a singular view, where only one subset is reflected can also be achieved by using the \textit{hide} button. For each filter set we provide three ways in which users can select groups of data points: \looseness=-1

\begin{enumerate}[wide ,nosep]
    \item[\textbf{1.}]  Model outcomes - a slider can be adjusted to establish the desired prediction confidence of the data points (Fig.~\ref{fig:interface} \textcircled{3}). This toggle is supplemented with a histogram showing the distribution of the points, colored according to their ground truth value. 
    
    \item[\textbf{2.}] Classification correctness – a clickable confusion matrix (Fig.~\ref{fig:interface} \textcircled{4}) can be accessed to create subgroups according to the correctness of the model's prediction (i.e., true positive, false negative).
    
    \item[\textbf{3.}] Data properties - points can be queried by setting a fixed value ranges for specific features (Fig.~\ref{fig:interface} \textcircled{5}). Multiple such filters can be aggregated to further narrow the search based on particular feature range requirements. 
\end{enumerate}


The visualization section acts to optimally represent the details and context of the data subsets isolated using the filtering options \textbf{(S2)}. Furthermore, counterfactual explanations are used to provide direct information about the decision space of the model \textbf{(S3)}. 

For the entire interface we use a format that splits the space first by feature column and then juxtaposes the two subsets being considered as sub-columns (Fig.~\ref{fig:interface} \textcircled{6}). In other words, the left \textit{Filter Set 1} corresponds to the left sub-column and the right \textit{Filter Set 2} to the right one. This conforms to our overarching premise that information can be best inferred from comparison. 


The contents of the sub-columns for each feature are largely customizable based on the level and depth of analysis desired by the user. As such, buttons can be used to alternate between varying degrees of detail shown about the distribution of the groups of data points. Three levels of detail are available - median, histogram, and individual data points - with each subsequent option providing more specific information about the distribution than the last (Fig.~\ref{fig:interface} \textcircled{7}). By default, only the first two toggles are activated as the visualization features ticks noting the median levels of each feature  (Fig.~\ref{fig:interface} \textcircled{8}) and histograms indicating the grouped distribution of the points (Fig.~\ref{fig:interface} \textcircled{9}). The histograms are constructed to reflect the binning parameters used to discretize the data in the counterfactual algorithm so that they provide transparency and consistency into the method. The final available toggle reveals individual points in the visualization. This acts to comprehensively characterize the data as the entire distribution can be fully observed. However, due to the feature based comparison approach adopted in the interface, the relationships across columns are not immediately obvious. Clicking on any of the single points reveals a set of parallel coordinates for all the data points in the same bin as the selection. 


In addition, to the distribution information, arrows are aggregated to signify the counterfactual explanations (Fig.~\ref{fig:interface} \textcircled{10}) for all those data points that fulfill the conditions set by the algorithm. The curved arrows indicate the necessary changes between histogram bins. The color of the counterfactual stems from the binary classification decision made by the model. For a \textit{positive} prediction, red arrows are used to show what changes would result in the decision becoming \textit{negative}, while green arrows are used for \textit{negative} instances as indicators for a \textit{positive} change. Opacity and perturbations help the user intuitively determine the amount of counterfactuals in a given feature column. Furthermore, a hovering action can be used to go beyond a singular analysis of features and highlight a complete counterfactual. This functionality is paramount as most explanations are comprised of concurrent expected changes in multiple features to reflect the complex pathways through which decisions are made. 

The number of customizable functions available within \textit{\app} can result in difficulty discerning between the important and redundant information. To alleviate this problem a multi-option sorting mechanism is built into the tool to reorder the vertical axes (Fig.~\ref{fig:interface} \textcircled{11}). Also, by default the tool separates the continuous and categorical features for a more natural separation.


While the application relies on visual cues to convey information, the quantitative information embedded in the interface can also be accessed directly using the details toggle. When switched on, the ranges of the features are depicted through textual labels. Furthermore, hovering over the histogram bins reveals the value range of the data points contained in that section.



\section{Use Case}

A detailed use case is used to evaluate \textit{\app} and demonstrate how it can be utilized in a specific context. The following is one example from stress testing our tool in several common scenarios, and as such provides a small overview of the tool's capabilities and limitations. \looseness=-1

The scenario described features the Home Equity Line of Credit (HELOC) dataset which was released as part of the FICO xML challenge ~\cite{fico}. It consists of applications made by homeowners in an attempt to get a credit line from the bank. The model aims to predict the binary variable \emph{Risk Performance} where \emph{bad} indicates that a consumer was at least 90 days past due once and \emph{good} that they always paid on time. In the financial sector, data assisted decision making such as this could help mitigate risk and quickly pinpoint individuals who might struggle to meet payment deadlines. \looseness=-1


This can be considered as an example of an \textbf{exploration} task where no concrete initial hypothesis can be made with regards to the decisions. Instead we envision the \textit{model developers} approaching this model with a plan to stress test it and identify any potential instances of poor performance.\looseness=-1

Starting the analysis by simply juxtaposing all the positive and negative datapoints helps get a sense of what the model considers a \emph{good} and \emph{bad} instance \textbf{(S1)}. Sorting by median instantly highlights which features have the greatest differences in this comparison \textbf{(S2)}. Since this dataset has in excess of 20 features, effective use of the sorting function is key to reading the visualization. \looseness=-1

\begin{figure}[t]
  \centering
  \includegraphics[width=1\linewidth]{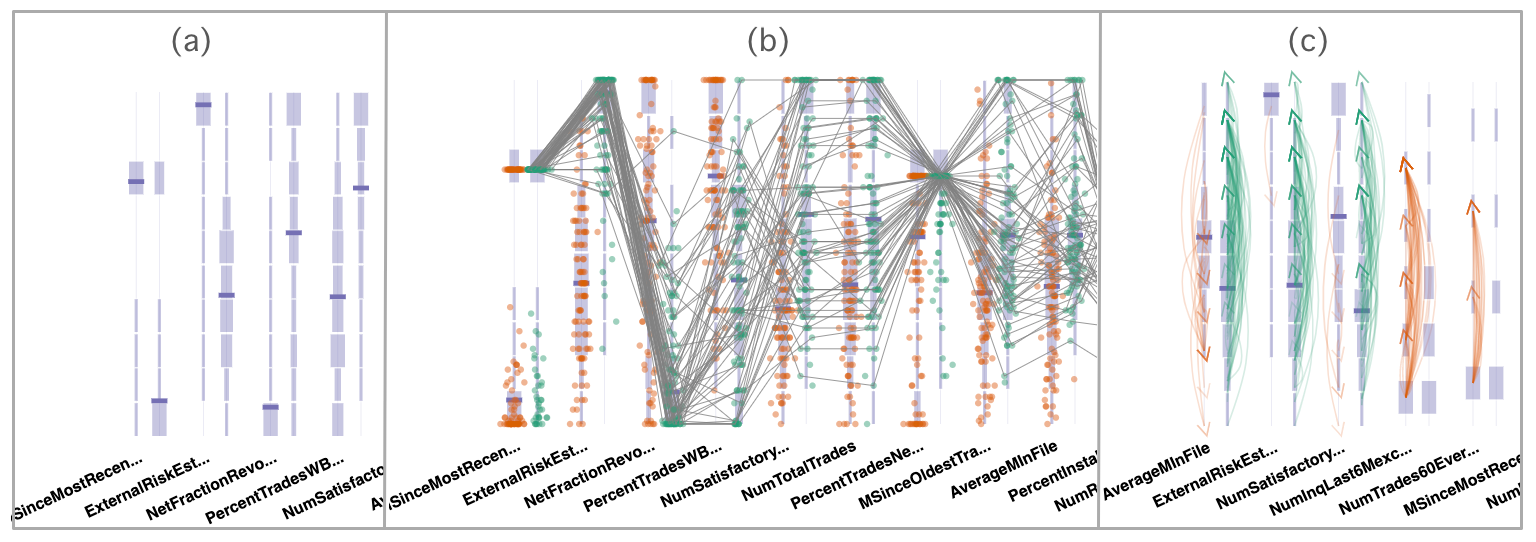}
   \vspace*{-.3cm}
   \caption{(a) Comparing distributions (b) Using parallel coordinates to observe trends (c) Highlighting common counterfactuals}
  \label{fig:UC2}
   \vspace*{-.5cm}
\end{figure}

The divergence of distribution is most visible in the features \emph{Months Since Last Delinquency}, \emph{External Risk Estimate} and \emph{Net Fraction Revolving Burden}. Immediately it becomes apparent that there is a single histogram bin that holds the majority of the positively categorized points in all of these aforementioned features (Fig.~\ref{fig:UC2} (a)). Using the parallel coordinates view (Fig.~\ref{fig:UC2} (b)) confirms an almost uniform trend where every positively predicted point follows the same path in these feature columns \textbf{(S3)}. This demonstrates the usefulness of this added functionality as parallel trends are very easily identifiable. \looseness=-1

To confirm consistency in the model's decision making, a new comparison between the True Positive (TP) and False Positive (FP) subsets, can be created \textbf{(S1)}. The distributions are now reflective of almost identical cohorts and the reason for the miss-classification is not obvious \textbf{(S2)}. As a starting point for further analysis the three properties deemed important in the previous comparison are examined. The feature range selector was utilized to identify that the clearest difference was in the \emph{External Risk Estimate} column where the bin range 80-82 continued to be densely populated for TP cases ($>50\%$ of points) while only a sample of the FP subset was contained in that range ($<20\%$ of points). Based on this contextual analysis we know this feature plays a key role in driving the decisions \textbf{(S3)}.\looseness=-1


To understand the model's internal logic we examine the counterfactual explanations (Fig.~\ref{fig:UC2} (c)). Reverting to the original positive versus negative comparison reaffirms the importance of \emph{External Risk Estimate} as a being weighted heavily by the model. More importantly, it reports \emph{Average Months in File} as being an essential component of almost all counterfactuals generated \textbf{(S3)}. In the context of the dataset and the model's purpose in real life applications this can be deemed an issue. The two key drivers of decisions are both time dependent variables which a potential client cannot immediately improve. Similarly, such a rigid model that lacks actionable features does not effectively assess the applicant's entire profile. \looseness=-1

This scenario also exemplifies a shortcoming in our \textit{\app} interface. It would be very useful to be able to limit how the counterfactuals explanations are generated in real time. For example, while the algorithm selects changes in the non immediately actionable features \emph{Average Months in File} and \emph{External Risk Estimate} as the quickest way to flip the decision, other pathways which make changes only in actionable features might also be possible.\looseness=-1

It should also be noted that not only are the visualized explanations concentrated in a select number of columns, there also appears to be inconsistencies depending on the classification outcome. The counterfactuals seen for the positive and negative cases do not depict opposite sets of changes. While this in itself does not represent flawed model behaviour, it does raise some concerns regarding its decision mechanisms. Studying the suggested changes for negatively classified points reveals large increases in the \emph{Number of Inquiries in the Last 6 months} column. However, applying these changes would make the values for this feature extend beyond even the positively predicted cohort. This can either mean that the model has a skewed perception of a \emph{good} applicant or simply that the constraints for the counterfactual algorithm were too loose \textbf{(S3)}. \looseness=-1

This scenario showcases the cyclic nature of model validation and exploration. Our proposed workflow is used in a repetitive manner to tackle the questions that arise in the exploratory analysis. Due to the presence of a multitude of complex features in this dataset the number of paths the data scientist can take is endless. The use case is meant to show one way the user might approach this challenge.  \looseness=-1

\section{Limitations \& Future Research} 


\textit{\app} acts as a great first step in assisting \textit{model developers} with model validation, however, we also acknowledge certain limitations. Firstly, scalability remains a key issue. Since counterfactuals need to be generated for the entire dataset, the computational requirements are proportional to input size. One solution would be to extend the random sampling method that is currently used within the visualization for optimization. A similar issue arises with high dimensional data as the visualization can effectively display only up to 30 features. However, this shortcoming can be mitigated through an active use of the sorting functionality which identifies the features worth visualizing. Furthermore, the tool is currently built for binary classification and tabular data, thus lacking generalizability. An extension to multi-class models should be straight forward through a target class selection option in the interface. However, introducing other types of input data such as images might require new solutions.

Future work will look to introduce the comparison of different models in addition to analysing different cohorts of data. In situations where two models have similar predictive performance, such an extension would be very useful in selecting the model that best coincides with the expected behaviour of the data scientist. While currently the user of \textit{\app} is tasked with finding patterns from the counterfactual explanations themselves, future iterations will look to systematically highlight trends and provide recommendations.
 

\section{Conclusion}

In this paper, we presented \textit{\app} – a tool aimed to guide and assist \textit{model developers} in the building and validation of black-box models. Through a deconstruction of common challenges faced by data scientists we proposed a novel workflow that can be used to facilitate a successful validation process. \textit{\app} caters to this framework and assists the user through a comparison based visual interface. Furthermore, our contribution rests in a novel algorithm to generate counterfactual explanations and visualize them in an aggregative manner.



\bibliographystyle{abbrv-doi}

\bibliography{main}
\end{document}